\documentclass[12pt,a4paper,final]{iopart}
\usepackage{graphicx}
\usepackage{iopams}
\usepackage{cite}
\usepackage[colorlinks=true,allcolors=blue,breaklinks=true]{hyperref}
\usepackage{cases}
\usepackage{bm}
\usepackage{color}

\newcommand*\chem[1]{\ensuremath{\mathrm{#1}}}

\begin{document}

\title[]{{Tuning Ginzburg-Landau theory to quantitatively study thin ferromagnetic materials}}

\author{Pamela C Guruciaga,$^1$ Nirvana Caballero,$^2$ Vincent Jeudy,$^3$ Javier Curiale,$^{4,5}$ Sebastian Bustingorry$^4$}
\address{$^1$ Centro At\'omico Bariloche, Comisi\'on Nacional de Energ\'{\i}a At\'omica (CNEA), Consejo Nacional de Investigaciones Cient\'{\i}ficas y T\'ecnicas (CONICET), Av. E. Bustillo 9500, R8402AGP San Carlos de Bariloche, R\'{\i}o Negro, Argentina}
\address{$^2$ Department of Quantum Matter Physics, University of Geneva, 24 Quai Ernest-Ansermet, CH-1211 Geneva, Switzerland}
\address{$^3$ Universit\'e Paris-Saclay, CNRS,  Laboratoire de Physique des Solides, 91405 Orsay, France.}
\address{$^4$ Instituto de Nanociencia y Nanotecnolog\'{\i}a, CNEA-CONICET, Centro At\'omico Bariloche, Av.~E.~Bustillo 9500, R8402AGP San Carlos de Bariloche, R\'{\i}o Negro, Argentina}
\address{$^5$ Instituto Balseiro, Universidad Nacional de Cuyo - CNEA, Av.~E.~Bustillo 9500, R8402AGP San Carlos de Bariloche, R\'{\i}o Negro, Argentina}

\ead{pamela.guruciaga@cab.cnea.gov.ar}


\begin{abstract}
    Along with experiments, numerical simulations are key to gaining insight into the underlying mechanisms governing domain wall motion in thin ferromagnetic systems. However, a direct comparison between numerical simulation of model systems and experimental results still represents a great challenge. Here, we present a tuned Ginzburg-Landau model to quantitatively study the dynamics of domain walls in quasi two-dimensional ferromagnetic systems with perpendicular magnetic anisotropy. This model incorporates material and experimental parameters and the micromagnetic prescription for thermal fluctuations, allowing us to perform material-specific simulations and at the same time recover universal features. We show that our model quantitatively reproduces previous experimental velocity-field data in the archetypal perpendicular magnetic anisotropy Pt/Co/Pt ultra-thin films in the three dynamical regimes of domain wall motion (creep, depinning and flow). In addition, we present a statistical analysis of the domain wall width parameter, showing that our model can provide detailed nano-scale information while retaining the complex behavior of a statistical disordered model.
\end{abstract}

\noindent{\it Keywords\/}: {interfaces in random media, dynamical processes, defects, numerical simulations}



\section{Introduction}\label{sec:intro}

Elastic driven interfaces are ubiquitous in nature. They appear in systems as diverse as contact lines in wetting~\cite{degennes1985wetting,ledoussal_contact_line}, earthquakes~\cite{jagla2010mechanical}, vortices in type-II superconductors~\cite{blatter_vortex_review}, and domain walls in ferroelectric~\cite{paruch_ferro_roughness_dipolar,paruch_ferro_quench,paruch2006nanoscale}, {ferrimagnetic}~\cite{caretta2018fast} and  ferromagnetic~\cite{lemerle_domainwall_creep,krusin_pinning_wall_magnet} materials. The latter, particularly, present very promising technological applications~\cite{Stamps2014} related to the possibility of tuning domain wall motion with controllable parameters such as electric currents or magnetic fields. 
The behavior of the velocity in these systems, however, depends not only on these external parameters but also on the relevant interactions between magnetic moments, 
the pinning disorder of the material, and thermal activation.
{Having a} theoretical approach that accounts for the interplay of all these ingredients and allows for {{a direct comparison between simulations and experiments}} {is of great importance} for the design and engineering of {magnetic} materials.

A general theory for the velocity-field response in {magnetic systems} was derived in the context of the elastic-line model~\cite{ioffe_creep,nattermann_creep_law,chauve_creep_long,bustingorry_thermal_rounding_epl,kolton_dep_zeroT_long,Ferrero2013,pardo2017,jeudy2018pinning}. {Within this formalism, the domain wall is modeled as an elastic interface by considering solely its position; any internal structure is neglected, thus ignoring any possible dynamics (e.g. precession) of the magnetic moments that compose it.} 
While in a perfectly ordered system the velocity-field relation is simply linear, quenched disorder is responsible for a much more complex behavior. 
When considering an applied external magnetic field $H$, and in the presence of quenched disorder, there is a field $H_{\mathrm{d}}$ --called the depinning field-- below which no domain wall motion can exist at $T=0$. 
At finite temperature, a thermally activated \emph{creep} regime appears at low fields following the law $\ln(v)\propto H^{-\mu}$, where the universal creep exponent is given by $\mu=\left({2\zeta_{\mathrm{eq}}+d-2}\right)/\left({2-\zeta_{\mathrm{eq}}}\right)$ with $d$ the dimension of the interface and $\zeta_{\mathrm{eq}}$ the equilibrium roughness exponent. In a one-dimensional interface, then, the theoretical value $\zeta_{\mathrm{eq}}=2/3$~\cite{huse1985pinning,kardar1985roughening} implies a creep exponent $\mu=1/4$, which has been found experimentally as well~\cite{lemerle_domainwall_creep,metaxas2007creep,ferre2013universal,caballero2017excess,savero2019universal,domenichini2019transient}.
In the \emph{depinning} regime, just above $H_{\mathrm{d}}$, the velocity presents universal power-law behavior associated to the underlying $T=0$ transition. Finally, $v\propto H$ at higher fields, in what is known as the \emph{flow} regime. The proportionality constant in this linear behavior is the mobility, and is the same than in the system without quenched disorder.

From the standpoint of statistical mechanics, a suitable model to study magnetic domains is given by Ginzburg-Landau theory, {with a proper inclusion of the disorder of the media}~\cite{jagla2004,jagla2005,caballero2018magnetic}. This theory was originally proposed as a mean-field approach to continuous phase transitions, relating the order parameter to the underlying symmetries of the system. Moreover, simple dissipative dynamics for the order parameter can be considered, allowing to study time-dependent phenomena~\cite{chaikin}. {When considering disorder, the obtained}  domain wall dynamics using this scalar-field {approach} presents the same non-linear response (creep, depinning and flow) for the velocity-field curve {as observed in experiments}~\cite{caballero2018magnetic}.

Models such as the aforementioned elastic line and Ginzburg-Landau theory have proven useful to understand the universal characteristics of domain wall dynamics~\cite{caballero2018magnetic}. {In particular, experiments and theoretical predictions are in good agreement for the values of the critical exponents $\mu$, $\beta$ and $\psi$, the two latter characterizing the velocity in the depinning regime [$v(H, T \to 0) \sim (H-H_{\mathrm{d}})^\beta$ and $v(H_{\mathrm{d}}, T)\sim T^\psi$].}
{However, these statistical models miss {material}-dependent characteristics, thus falling short in the connection to experiments.}
Micromagnetic theory, on its turn, provides important insight into the physical properties of magnetic materials, with particular care of material and experimental parameters~\cite{Lopez-Diaz2012,boulle2013domain,vansteenkiste2014design,voto2016disorder,pfeiffer2017geometrical,leliaert2018fast}.
{Although micromagnetic simulations are particularly useful to describe the dynamics of magnetic textures in flow regimes at zero temperature, taking into account the contributions of domain wall pinning and thermal fluctuations to obtain critical behavior is not straightforward.}

\begin{figure}[tb]
    \centering
    \includegraphics[width=0.7\linewidth]{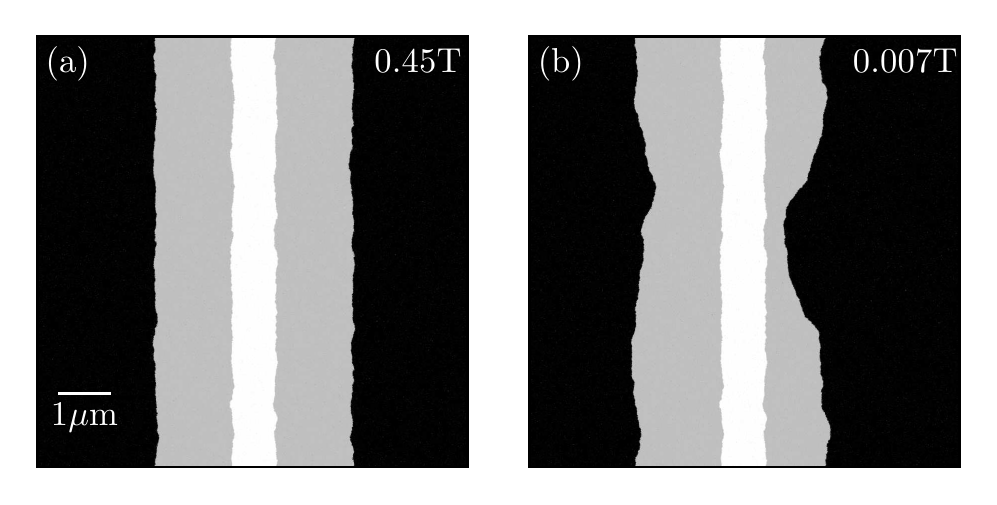}
    \caption{Spatial distribution of the out-of-plane magnetization $m_z$ for a two-dimensional system with quenched Gaussian Voronoi disorder ($\epsilon=0.19$ and $\ell=30\,\mathrm{nm}$, {see \Sref{sec:validation}}) at $T=300\,\mathrm{K}$ before and after applying an out-of-plane magnetic-field pulse of (a)~$\mu_0H_z=0.45\,\mathrm{T}$ during $\delta t=12\,\mathrm{ns}$, and (b)~$\mu_0H_z=0.007\,\mathrm{T}$ during $\delta t=10\,\mu\mathrm{s}$. While the white strip corresponds to the initial relaxed domain with $m_z=+1$, gray represents its growth after the pulse. Black stands for $m_z=-1$.}
    \label{fig:system2}
\end{figure}

In this work, {we present a connection between micromagnetism and Ginzburg-Landau theory that allows us to quantitatively study ferromagnetic ultra-thin films by means of a tuned scalar-field model. In particular, this material-dependent approach incorporates thermal fluctuations following the micromagnetic prescription, which results in a non-trivial noise term.}
{{As shown in \fref{fig:system2}, we are able to simulate the growth of magnetic domains and the concomitant domain wall dynamics, which can be directly compared with polar magneto-optical Kerr effect (PMOKE) experiments}} (we will return to \fref{fig:system2} in \Sref{sec:validation}).
{We use previous experimental velocity-field data in \chem{Pt/Co/Pt} ultra-thin films~\cite{metaxas2007creep} to test our material-dependent model. Not only we find good agreement in the three regimes of domain wall motion, but we also recover universal features as the creep exponent. In addition, we report new results regarding domain wall width fluctuations in this system, which are typically not exposed by PMOKE experiments.} 
{In this way, our tuned Ginzburg-Landau model shows great versatility to perform numerical simulations of domain wall dynamics in ultra-thin magnetic materials with perpendicular magnetic anisotropy.}

\section{From micromagnetism to the Ginzburg-Landau model}\label{model}

\subsection{Ginzburg-Landau theory for magnetic systems}\label{sec:GL}

{As mentioned in the previous section, {from a statistical physics standpoint,} Ginzburg-Landau theory can be used to study magnetic systems by relying on the symmetries of the problem and considering the interactions via effective quantities. Since we are interested in studying the case of a ferromagnetic thin film with perpendicular easy-axis anisotropy, the non-conserved scalar field $\phi(\bi{r},t)$ is taken to represent the out-of-plane component of the magnetization. In the limit of strong perpendicular anisotropy and strong damping~\cite{nicolao2007}, this generalized order parameter follows the evolution equation
\begin{equation}
    {\partial_{t} \phi} = -\Gamma\frac{\delta\mathcal H_{\mathrm{GL}}}{\delta\phi} + \xi \, .
\end{equation}
Here, ${\Gamma}$ is a damping parameter that sets the time scale of the problem, and $\xi(\bi{r},t)$ is an additive, uncorrelated white noise that represents thermal fluctuations and satisfies
\begin{equation}
    \langle \xi(\bi{r},t) \rangle =0 
    \quad \mathrm{and} \quad
    \langle \xi(\bi{r},t)\xi(\bi{r}',t')\rangle =2\Gamma T\delta(\bi{r}-\bi{r}')\delta(t-t') \, ,
\end{equation}
{where $T$ is the temperature}~\cite{chaikin}. As discussed in~\cite{jagla2004,jagla2005,caballero2018magnetic}, the free-energy Hamiltonian $\mathcal H_{\mathrm{GL}}$ can be modeled by following the modified ``$\phi^4$ model'' and consists of three contributions: 
\begin{equation}
    \mathcal{H}_{\mathrm{GL}} = \beta \int \frac{\left| \nabla\phi \right|^2}{2} \rmd\bi{r} + \alpha \int \left(-\frac{\phi^2}{2}+\frac{\phi^4}{4}\right)\rmd\bi{r} - h \int\phi\, \rmd\bi{r} \, .
\end{equation}
While the first term incorporates the rigidity of the system with elastic stiffness $\beta$, the second one represents easy-axis anisotropy (favoring the values $\phi=\pm 1$) by means of a two-well potential with an energy barrier $\alpha$.
The last term, on its turn, amounts to the inclusion of an external magnetic field $h$ perpendicular to the film.

{Although the range of $\phi$ does not need to be bounded \textit{per se}}, its physical interpretation as a component of the magnetization implies that $|\phi|\leq 1$. In order to promote this, {a term $h \phi^3/3$ {can} be subtracted from the external field contribution, resulting in a term proportional to $h(1-\phi^2)$ in the evolution equation~\cite{jagla2005}}.
Taking all these ingredients into account, the Langevin equation for the modified Ginzburg-Landau scalar-field model is
\begin{equation}\label{phi4}
    (1/{\Gamma})\, \partial_t\phi = \beta\,\nabla^2\phi +  \left(\alpha\,\phi + h\right)\left(1-\phi^2\right) + \tilde\xi \, ,
\end{equation}
with $\tilde\xi = \xi/\Gamma$. As shown in~\cite{chaikin}, {the equilibrium value of $\phi$} yields domain walls with a width {parameter} given by $\sqrt{{2\beta}/{\alpha}}$ and a domain wall energy equal to $({4}/{3})\sqrt{2\beta \alpha}$. These results will be useful in \Sref{sec:link}, where we establish the connection between this scalar-field model and micromagnetism.

With some variations, this model has been widely used to {generically} model the magnetization in quasi two-dimensional systems~\cite{jagla2004,jagla2005,caballero2018magnetic,nicolao2007,PerezJunquera2008,marconi2011crossed}. Nonetheless, the lack of material and experimental parameters {make it unsuitable for quantitative comparisons with real systems.}
}

\subsection{Micromagnetic theory}
\label{sec:mmtheory}

{In this section, we present the micromagnetic approach to the dynamics of a ferromagnetic thin film with perpendicular magnetic anisotropy.}
A ferromagnet is a system that presents a net magnetization (that is, magnetic moment per unit volume) in the absence of external field. It can be thought of as composed by cells of a given volume which is large compared to the atomic scale, and containing a great number of atomic magnetic moments. Then, micromagnetism provides a way to model the evolution of the magnetization $\bi{M}(\bi{r},t)$ in each cell by following the stochastic Landau-Lifshitz-Gilbert (SLLG) equation. In the Landau formulation (see~\cite{roma2014numerical} and references therein) it is given by
\begin{equation}\label{SLLG}
    \partial_t\bi{M}
    =-\frac{{\gamma}\mu_0}{1+\eta_0^2} \, \bi{M}\times \left[\left(\bi{H}_{\mathrm{eff}} +\bi{H}_{\mathrm{th}}\right) \vphantom{\frac{\eta_0}{M_{\mathrm{S}}}}\right. \left. +\frac{\eta_0}{M_{\mathrm{S}}}\bi{M}\times\left(\bi{H}_{\mathrm{eff}}+\bi{H}_{\mathrm{th}}\right)\right]\, ,
\end{equation}
where $\eta_0$ is the {adimensional} damping constant, $\mu_0$ is the vacuum permeability, {$\gamma$ is the gyromagnetic ratio} and $M_{\mathrm{S}}$ is the saturation magnetization. 
While the effective field is derived from the free-energy Hamiltonian $\mathcal{H}$ of the system by the relation
\begin{equation}\label{funcderiv}
    \mu_0\bi{H}_{\mathrm{eff}}=-\frac{\delta\mathcal{H}}{\delta\bi{M}}\, , 
\end{equation}
$\bi{H}_{\mathrm{th}}=f_x\hat{\bi{e}}_x+f_y\hat{\bi{e}}_y+f_z\hat{\bi{e}}_z$ is a random vector field that represents thermal noise. {The fluctuating terms of different cells are independent from each other, as well as the three components, and they are uncorrelated in time. The statistical properties of this vector white noise are then}
\begin{equation}
    \langle f_{\kappa i}(t)\rangle =0 
    \quad \mathrm{and} \quad
    \langle f_{\kappa i}(t)f_{\lambda j}(t')\rangle =2D\delta_{\kappa\lambda}\delta_{ij}\delta(t-t')
\end{equation}
for cells $\kappa$ and $\lambda$, Cartesian coordinates $i,j=x,y,z$, and 
\begin{equation} \label{TD}
    D=\frac{\eta_0k_BT}{{\gamma} V M_{\mathrm{S}} \mu_0^2}\, ,
\end{equation}
with $k_B$ the Boltzmann constant, $T$ the temperature of the system and $V$ the volume of the cell.\footnote{It is not straightforward to apply this methodology to the continuous theory of micromagnetism~\cite{Lopez-Diaz2012}. However, for a discrete formulation like ours, the thus defined random field has been shown to correctly reproduce equilibrium thermodynamics~\cite{brown1963thermal}.}

{In this work, the general micromagnetic formalism introduced above is adapted to study the magnetization dynamics in ultra-thin films with dominant uniaxial anisotropy. We are interested in studying the evolution of the $z$ component of the magnetization in a quasi two-dimensional system.} We consider exchange interactions with stiffness $A$, effective easy-axis anisotropy with strength $K$, and Zeeman coupling to an out-of-plane magnetic field of intensity $H_z$.
The free energy of the system is given by
\begin{equation}\label{energyH}
    \mathcal{H}= A \int \big|\nabla\bi{m}(\bi{r})\big|^2 \rmd\bi{r} - K \int \left[\hat{\bi{e}}_z\cdot\bi{m}(\bi{r})\right]^2 \rmd\bi{r} -\mu_0 H_z M_{\mathrm{S}} \int \hat{\bi{e}}_z\cdot \bi{m}(\bi{r}) \rmd\bi{r}
    \, ,
\end{equation}
where $\bi{m}=\bi{M}/M_{\mathrm{S}} = m_x \hat{\bi{e}}_x + m_y \hat{\bi{e}}_y + m_z \hat{\bi{e}}_z$ (with the norm constraint $|\bi{m}|=1$).

Following the prescription of equation~\eref{funcderiv} to calculate the effective field, equation~\eref{SLLG} for the $z$ component of the magnetization can be written
\begin{eqnarray}\label{dtdmz}
    \frac{1+\eta_0^2}{ {\gamma}\mu_0\eta_0}\, \partial_t m_z= & \,
    a\,\nabla^2m_z + \left(k\, m_z+H_z\right)\left(1-m_z^2\right) \nonumber \\
     & + f_x \left({m_y}/\eta_0-{m_x}m_z\right) 
     + f_y  \left({m_x}/\eta_0-{m_y}m_z\right) \nonumber\\
     & + f_z \left(1-m_z^2\right) + {N} \, ,
\end{eqnarray}
where for simplicity we have defined $a=2A/M_{\mathrm{S}}$, $k=2K/M_{\mathrm{S}}$ {(known as the anisotropy field)}, and
\begin{equation}
    {N}= a m_z\left[({\nabla}{m_x})^2+({\nabla}{m_y})^2+({\nabla}m_z)^2\right]
    -\frac{a}{\eta_0}\left({m_x}\nabla^2 {m_y}-{m_y}\nabla^2 {m_x}\right) .
    \label{Nterm}
\end{equation}
The full micromagnetic description is comprised of the other two coupled equations for $m_x$ and $m_y$, analogous to equation~\eref{dtdmz}, and the norm constraint. 
Within this formalism, static domain wall solutions are described by domain wall width parameter equal to ${\Delta_0=}\sqrt{{A}/{K}}$ {(the width itself is $\pi\Delta_0$) and a domain wall energy given by $\varepsilon = 4\sqrt{AK}$}.\footnote{{Strictly, these expressions are valid when no dipolar coupling is considered, as in our case, or in a system with dipolar interactions and Bloch-type walls. Otherwise, a term proportional to the demagnetizing energy must be added to $K$ (see~\cite{Malozemoff}).}}

\subsection{Linking the models}\label{sec:link}

\begin{figure}[tb]
    \centering
    \includegraphics[width=0.7\linewidth]{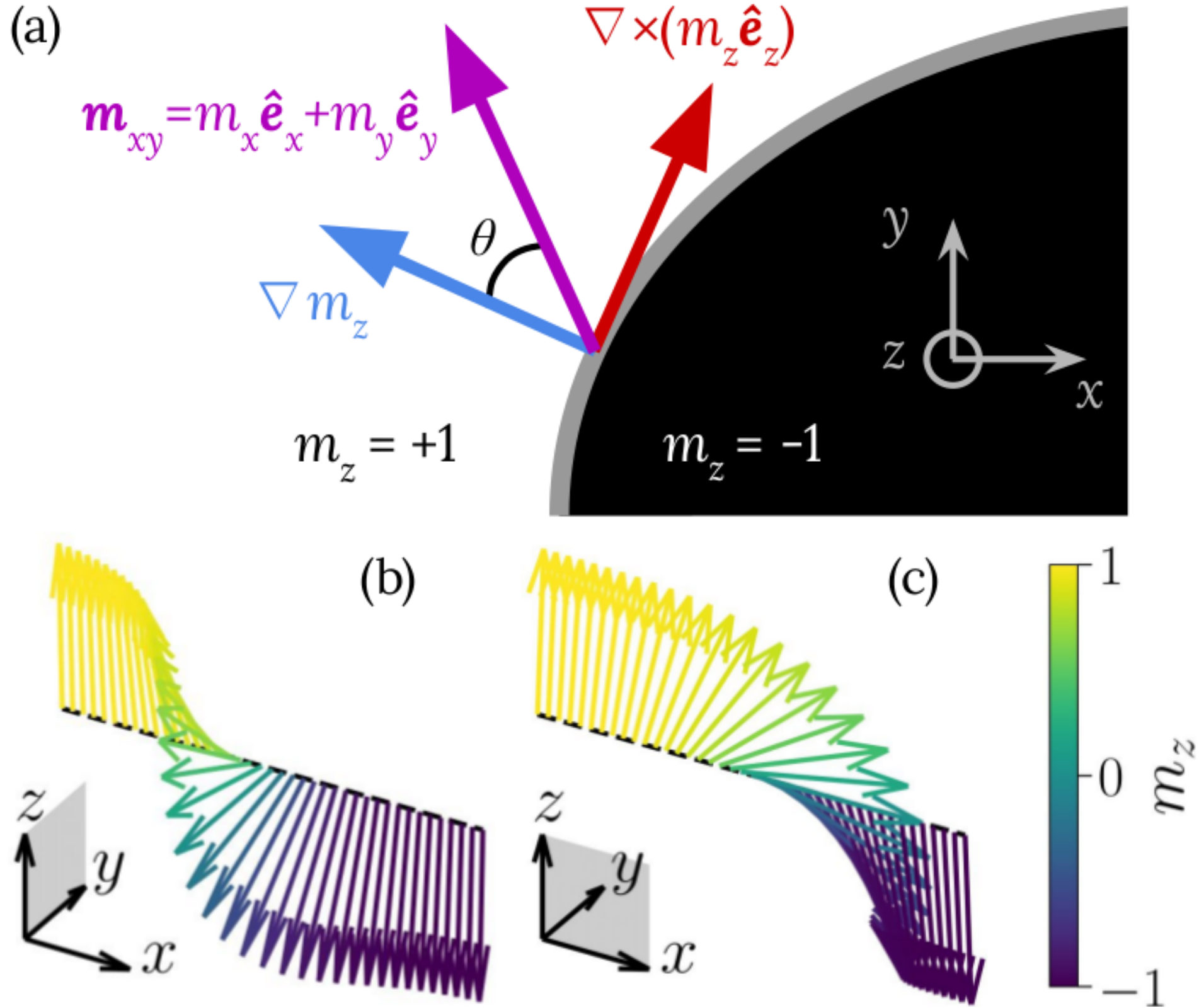}
    \caption{Local parametrization of the in-plane magnetization. (a)~Fixing the angle $\theta$ allows writing $\bi{m}_{\mathrm{xy}}$ in terms of components parallel and perpendicular to $\nabla m_z$. The choice of $\theta$ determines the kind of domain walls present in the system, two particular cases being (b)~Bloch and (c)~N\'eel walls. In these configurations, $\bi{m}$ (arrows) rotates perpendicularly to $\nabla m_z$ and $\nabla \times (m_z\hat{\bi{e}}_z)$, respectively, and is contained in the corresponding gray planes.}
    \label{fig:diagmxy}
\end{figure}

Our aim here is to obtain an effective description in terms of $m_z$ using one single evolution equation. This will represent a loss of generality of the formalism {presented in \Sref{sec:mmtheory}, but it will allow us to link the micromagnetic theory with the Ginzburg-Landau statistical model of \Sref{sec:GL}.}
{To achieve this, we uncouple equation~\eref{dtdmz} from the ones for the other components of $\bi{m}$ by writing $\bi{m}_{\mathrm{xy}}=m_x \hat{\bi{e}}_x + m_y \hat{\bi{e}}_y$ as a function of $m_z\equiv m_z(x,y)$. As shown in} \fref{fig:diagmxy}(a), {we {do so} by writing}
\begin{equation}\label{mxy}
    \bi{m}_{\mathrm{xy}}=\sqrt{1-m_z^2}\, \left[ \cos\theta \frac{\nabla m_z}{|\nabla m_z|} +\sin\theta\frac{\nabla\times (m_z\hat{\bi{e}}_z)}{|\nabla \times (m_z\hat{\bi{e}}_z)|}\right]
\end{equation}
{with a constant $\theta$, thus locally fixing the direction of the in-plane magnetization.}
In this way, the in-plane part of $\bi{m}$ is decomposed in one component in the direction of the gradient of $m_z$ and one component orthogonal to it.
Note that fixing the value of $\theta$ reduces the number of degrees of freedom so that only one evolution equation --in this case, that for $m_z$-- is needed to describe the magnetic properties of the system. The angle $\theta$ is a constant that generally relates $\bi{m}_{\mathrm{xy}}$ to $\nabla m_z$ in each cell.
Indeed, although we used a curved domain wall for the diagram in \fref{fig:diagmxy}(a), the proposed parametrization is actually independent of the existence of a domain wall in the system: any finite $\nabla m_z$ implies a finite $\bi{m}_{\mathrm{xy}}$.
{In the case where a domain wall is present, the angle between its normal and the in-plane magnetization in the center of the wall (where $m_z=0$) is equal to $\theta$. In this way, choosing the angle for the parametrization implies deciding \textit{a priori} if the wall will be Bloch ($\theta=\pi/2,\,3\pi/2$; see \fref{fig:diagmxy}(b)), N\'eel ($\theta=0,\, \pi$; see \fref{fig:diagmxy}(c)) or anything in between, and its chirality.}

{Equation~\eref{mxy} is used to parameterize the magnetization in terms of an angle $\theta$ that is not only constant along the wall but also does not change in time. As stated at the beginning of this section, this
represents a great simplification of the SLLG equation. At variance with the elastic-line model discussed in \Sref{sec:intro}, the micromagnetic formulation presents non-trivial behavior even in the absence of quenched disorder, with the velocity-filed response of the domain wall coupled to the dynamics of the magnetization orientation inside the domain wall~\cite{Malozemoff}.
For a driving field lower than the so-called Walker field ($H < H_{\mathrm{W}}$), the orientation of the magnetization within the domain wall remains stationary and the velocity follows a linear regime with $v \propto 1/\eta_0$. On the contrary, above $H_{\mathrm{W}}$ magnetization inside the domain wall precesses during the motion. The velocity initially decreases with increasing drive and then reaches an asymptotic linear regime for $H\gg H_{\mathrm{W}}$ with $v \propto \eta_0/(1+\eta_0^2)$. For most experiments (as those reported in~\cite{metaxas2007creep}), where disorder must also be considered, the depinning field $H_{\mathrm{d}}$ is significantly higher than $H_{\mathrm{W}}$, so above the creep and depinning regimes only the linear asymptotic precessional flow is observed.
At first sight, the latter could not be modeled with a fixed $\theta$. Nonetheless, as justified at the end of this section, our tuned Ginzburg-Landau approach will end up being independent of this angle, and thus able to effectively model the precessional regime.}

{The next step in relating micromagnetism and Ginzburg-Landau theory is to assess the significance of the term $N$ as defined in equation~\eref{Nterm}. If we set it to zero, it is easy to see that, at zero temperature, equation~\eref{dtdmz} becomes analogous to equation~\eref{phi4}. Following \Sref{sec:GL}, domain walls obtained in these conditions will have a larger width {parameter} given by $\sqrt{{2 A}/{K}}$ and a smaller domain wall energy equal to $({4}/{3})\sqrt{2 A K}$.}
Explicitly including {${N}$} in equation~\eref{dtdmz} can then be avoided if two new parameters $A_{\mathrm{eff}}$ and $K_{\mathrm{eff}}$ are conveniently defined in order to provide the {proper} micromagnetic domain wall width {parameter} and domain wall energy {mentioned at the end of \Sref{sec:mmtheory}}. 
The effective anisotropy and stiffness constants can be found by simply solving the system of equations~\cite{chaikin,Malozemoff}
\begin{numcases}{}
    \sqrt{\frac{2A_{\mathrm{eff}}}{K_{\mathrm{eff}}}}=\sqrt{\frac{A}{K}} \label{effectiveAK} \\
    \frac{4}{3}\sqrt{2A_{\mathrm{eff}}K_{\mathrm{eff}}}=4\sqrt{AK}
\end{numcases}
and thus obtaining $A_{\mathrm{eff}}= 3A/2$ and $K_{\mathrm{eff}}= 3K$.
Taking this into account, equation~\eref{dtdmz} becomes
\begin{eqnarray}\label{mzphi4}
    \frac{1+\eta_0^2}{ {\gamma}\mu_0\eta_0}\, \partial_t m_z= &\,
    a_{\mathrm{eff}}\,\nabla^2m_z + \left(k_{\mathrm{eff}}\, m_z+H_z\right)\left(1-m_z^2\right) \nonumber \\
    & + f_x \left({m_y(m_z)}/\eta_0-{m_x(m_z)}\,m_z\right) \nonumber \\
    & + f_y  \left({m_x(m_z)}/\eta_0-{m_y(m_z)}\,m_z\right) \nonumber \\
    & + f_z \left(1-m_z^2\right) \, ,
\end{eqnarray}
{with $a_{\mathrm{eff}}=2A_{\mathrm{eff}}/M_{\mathrm{S}}$ and $k_{\mathrm{eff}}=2K_{\mathrm{eff}}/M_{\mathrm{S}}$}. {Notice that we have written $m_x$ and $m_y$ in terms of $m_z$ via the parametrization presented in equation~\eref{mxy}.}

{As shown in \Sref{sec:mmtheory}, thermal fluctuations in micromagnetism are not included by means of a trivial additive white noise. In equation~\eref{mzphi4}, the multiplicative terms involving the random-field components $f_i$ make the noise different from zero only close to the domain wall, ensuring at the same time that $|m_z| \leq 1$.}
{Despite the different temperature implementation, {identifying $m_z\leftrightarrow\phi$ makes it clear that} evolution equation~\eref{mzphi4} has the shape of the well-known Ginzburg-Landau model for phase transitions presented in equation~\eref{phi4}, where the factors of the model can now be expressed as
\begin{equation}
    \Gamma = \frac{\gamma\mu_0\eta_0}{1+\eta_0^2} \, , \quad 
    \beta = \frac{3A}{M_{\mathrm{S}}} \, , \quad 
    \alpha = \frac{6K}{M_{\mathrm{S}}} \quad \mathrm{and} \quad 
    h = H_z  \, .
\end{equation}
Equation~\eref{mzphi4} then constitutes the tuned Ginzburg-Landau model. It incorporates material and laboratory parameters, as well as multiplicative thermal fluctuations. Considering the approximations made, it also represents a simplification of the micromagnetic model since it is written in terms of one single component of the magnetization.}

{Finally, given that our model does not include in-plane interactions of any kind (e.g. in-plane magnetic field, Dzyaloshinskii-Moriya interactions or dipolar coupling), it should be noticed that the magnetic moments have no preference for any particular value of the parametrization angle $\theta$. Indeed, this angle appears only in the temperature terms --which are basically random numbers--, yielding identical results for both N\'eel and Bloch walls.}
In the following section, numerical simulations of this model are compared with experimental results measured by Metaxas and collaborators~\cite{metaxas2007creep}.

\section{{Domain wall dynamics}}\label{sec:validation}

{We now compare velocity-field curves calculated with the {tuned Ginzburg-Landau} model and those obtained by PMOKE experiments. We solve equation~\eref{mzphi4} through a semi-implicit Euler scheme~\cite{caballero2018magnetic,ferrerophi4}, and perform simulations following the typical PMOKE experimental protocol (see for example~\cite{lemerle_domainwall_creep,metaxas2007creep,caballero2018magnetic}).}
The $L\times L\times s$ system of side $L$ and thickness $s$ is initialized with a narrow $m_z=+1$ stripe (white regions in \fref{fig:system2}(a) and \fref{fig:system2}(b)) surrounded by $m_z=-1$, and is allowed to relax {at zero magnetic field.}
Then, a positive field $H_z$ is applied, favoring the growth of the $+1$ domain until a certain maximum area of the growing domain is reached. At that point, the field is removed and the system is again allowed to relax. We calculate the domain wall velocity as
\begin{equation}
    v=\frac{1}{2L}\frac{\delta a}{\delta t}\, ,
\end{equation}
with $\delta a$ the difference in area between the final and initial relaxed configurations (gray regions in \fref{fig:system2}(a) and \fref{fig:system2}(b)) and $\delta t$ the duration of the field pulse.

\begin{figure}[tb]
    \centering
    \includegraphics[width=0.7\linewidth]{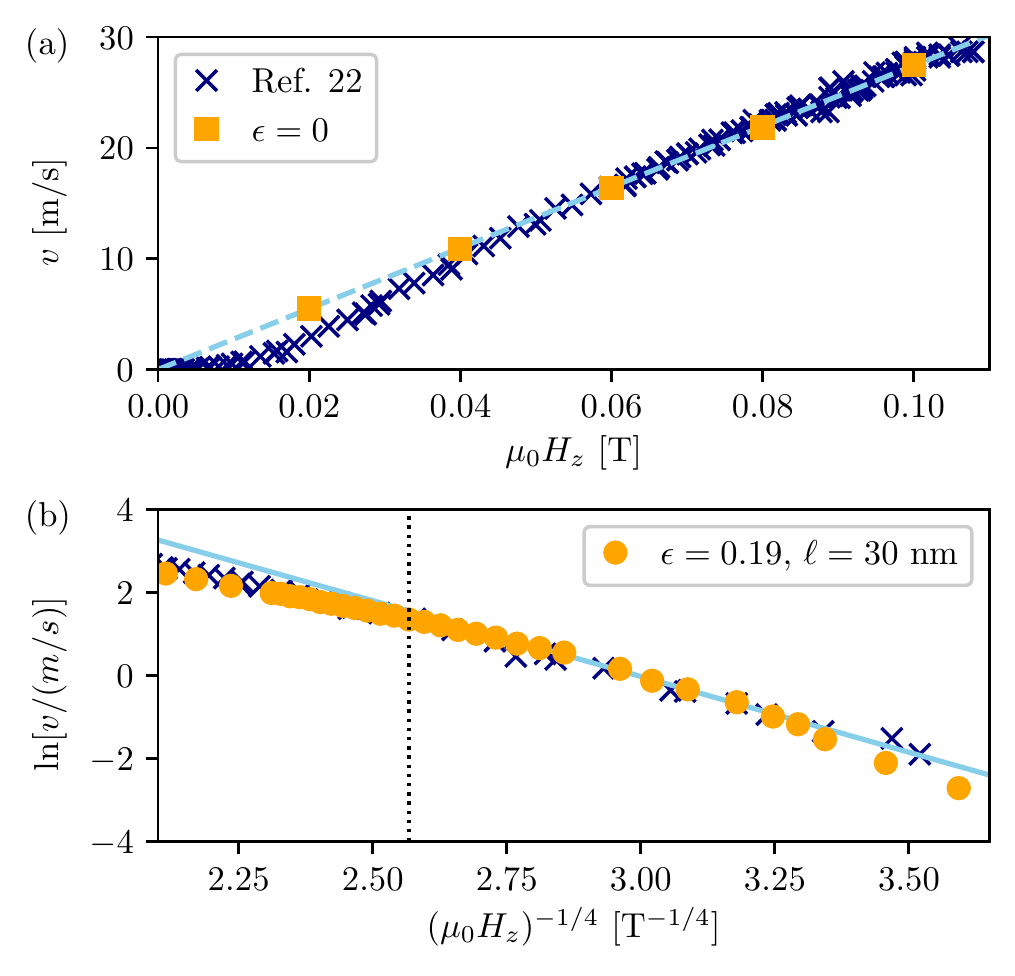}
    \caption{Comparison of our results (squares and circles) with experimental data (crosses) from~\cite{metaxas2007creep}. (a)~A system without quenched disorder recovers the expected velocity vs. magnetic field linear behavior, with a mobility consistent with the experimental value (dashed line). The only fitting parameter is the damping constant, which was found to be {$\eta_0=0.255$, {which is close to the value $0.27$} reported in~\cite{metaxas2007creep}}. (b)~The creep regime is also recovered by simulating a system with Gaussian Voronoi disorder characterized by $\epsilon=0.19$ and $\ell=30\,\mathrm{nm}$, at a temperature $T=300\,\mathrm{K}$. The {solid} line is a fit of the experimental data in the linear region of $\ln(v)$ vs. $(\mu_0H_z)^{-1/4}$, {while the dotted line shows the position of the depinning field $\mu_0 H_{\mathrm{d}} = 0.023\,\mathrm{T}$~\cite{metaxas2007creep} in the creep plot}.}
    \label{fig:exp-sim}
\end{figure}

The material parameters are taken to be those of Pt($4.5\,\mathrm{nm}$)/Co($0.5\,\mathrm{nm}$)/Pt($3.5\,\mathrm{nm}$) ultra-thin films at $T=300\,\mathrm{K}$, as reported by Metaxas \textit{et al.}~\cite{metaxas2007creep}: $A=1.4\times 10^{-11}\,\mathrm{J/m}$, $K=3.2\times 10^5\,\mathrm{J/m}^3$ and $M_{\mathrm{S}}=9.1\times 10^5\,\mathrm{A/m}$.
We work on a system of side $L=8.192\,\mu\mathrm{m}$ and thickness $s=0.5\,\mathrm{nm}$, with simulation cells of volume $V=l\times l\times s$ with $l=2\,\mathrm{nm}$.
{As shown in \fref{fig:exp-sim}(a), the damping constant is fitted in order to recover the flow regime observed in the experiment; we find $\eta_0=0.255$, i.e. within $6\%$ difference from {the value $0.27$ reported in~\cite{metaxas2007creep}}.}
This quantity is compatible with the precessional flow suggested in~\cite{metaxas2007creep} and recently confirmed in~\cite{herranen2019barkhausen}. 
{The reason for this might be related to the fact that, as no in-plane contributions to the free energy are being considered, the Walker field is $H_{\mathrm{W}}=0$~\cite{Malozemoff}.}
In this limit where $H_z\gg H_{\mathrm{W}}$ for any finite field, the model can be thought to effectively be in the precessional regime even though the angle between the domain wall and $\bi{m}_{\mathrm{xy}}$ is a constant. This fact is quite reasonable given that --as explained at the end of \Sref{sec:link}-- equation~\eref{mzphi4} does not depend on $\theta$.

\begin{figure}[tb] 
    \centering
    \includegraphics[width=0.7\linewidth]{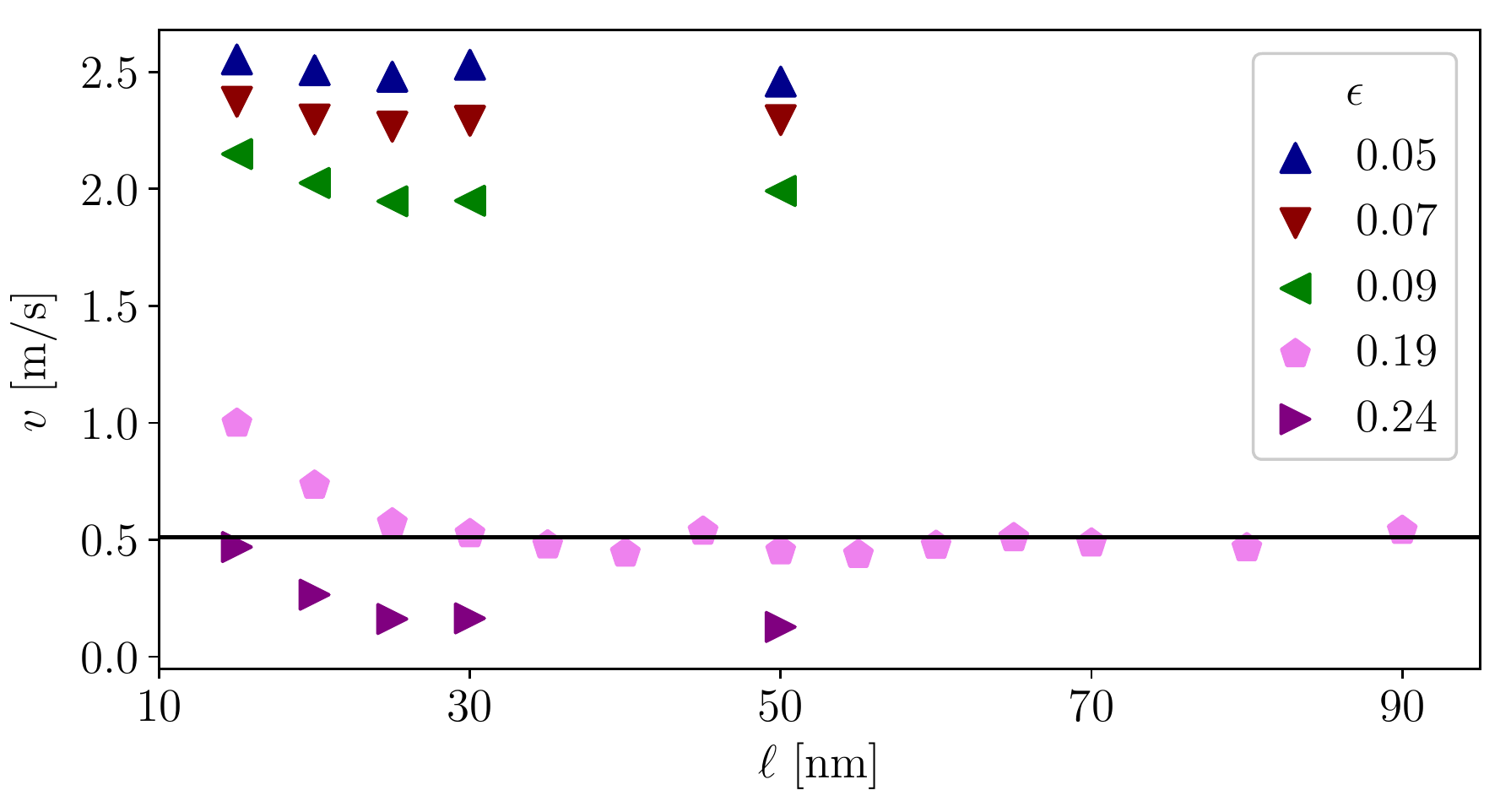}
    \caption{Different sets of parameters $\epsilon$, $\ell$ of the Gaussian Voronoi disorder are used in order to obtain the experimental value~\cite{metaxas2007creep} of the velocity (black full line) for a field $\mu_0H_z=0.01\,\mathrm{T}$ in the creep regime. We choose $\epsilon=0.19$, $\ell=30\,\mathrm{nm}$.}
    \label{fig:disorder}
\end{figure}

Afterwards, quenched disorder is included by means of a Voronoi tesselation with mean grain size $\ell$. The anisotropy is modified as $k_{\mathrm{eff}}\to  k_{\mathrm{eff}}\left[1+\epsilon\zeta(\bi{r})\right]$, where $\epsilon$ is the disorder intensity and $\zeta(\bi{r})$ has a constant value for each grain obtained from a Gaussian distribution {with zero mean and unit variance}~\cite{caballero2018magnetic}. {To determine the disorder parameters of the system, we {simulated domain-wall dynamics with} diverse values of $\epsilon$ and $\ell$ (\fref{fig:disorder}) for a field in the creep regime ($\mu_0H_z=0.01\,\mathrm{T}$). The resulting velocities are compared with the corresponding experimental value, and the best-fitting set of parameters is found. 
The best agreement with the experimental velocity is obtained with $\epsilon = 0.19$. For this value, there is a range for the Voronoi grain size, $\ell\geq 30\,\mathrm{nm}$, where the agreement is good. The Voronoi grain size can be associated with the {characteristic disorder length scale $\rho$} and it has been shown that typically $\Delta_0<\rho<L_{\mathrm{c}}$ (see~\cite{jeudy2018pinning}), where $L_{\mathrm{c}}$ is the Larkin length. Given that in our system $L_{\mathrm{c}} \approx 40\,\mathrm{nm}$~\cite{jeudy2018pinning} {(and $\Delta_0=6.6\,\mathrm{nm}$)}, in the following we shall use {$\ell = 30\,\mathrm{nm}\approx\rho$}.} {The disorder intensity, on its turn, represents the width of the anisotropy distribution, which is not known \textit{a priori}. Thus, fitting the disorder parameters as described above in a family of samples would allow for a very interesting characterization of disorder in real materials.}

{We have so far tuned the missing parameters of the problem to represent \chem{Pt/Co/Pt} thin films. The value of the damping was found by comparing simulations of a system without disorder with the experimental flow regime. The disorder, on its turn, was characterized using only one point of the experimental velocity-field curve in the creep regime. As shown in \fref{fig:exp-sim}(b), and in a remarkable display of robustness, simulations with the aforementioned parameters allow us to quantitatively reproduce experimental $v$ vs. $H_z$ data~\cite{metaxas2007creep} in the \emph{three} regimes: creep, depinning and flow. This validates} the use of the {proposed model} to obtain domain wall velocities {for systems with strong perpendicular magnetic anisotropy}.
{Although} our simulations are material-dependent, universal features like the creep exponent $\mu$ are clearly recovered. Also, {in spite of being a scalar-field version of the SLLG equations}, our model is the first to quantitatively reproduce these experimental results. {Indeed, a previous attempt to fit the same data can be found in~\cite{herranen2019barkhausen}, but full micromagnetic simulations were performed at $T=0$ only~\cite{herranen_thesis}.
In~\cite{shahbazi2018magnetic}, on its turn, finite-temperature numerical simulations of a micromagnetic model were used to study the velocity-field response in a related family of materials (\chem{Pt/Co/Au_xPt_{1-x}}). Although no direct comparison with the creep regime was made, results present good agreement in the depinning and flow regimes.}

\begin{figure}[t!]
    \centering
    \includegraphics[width=0.7\linewidth]{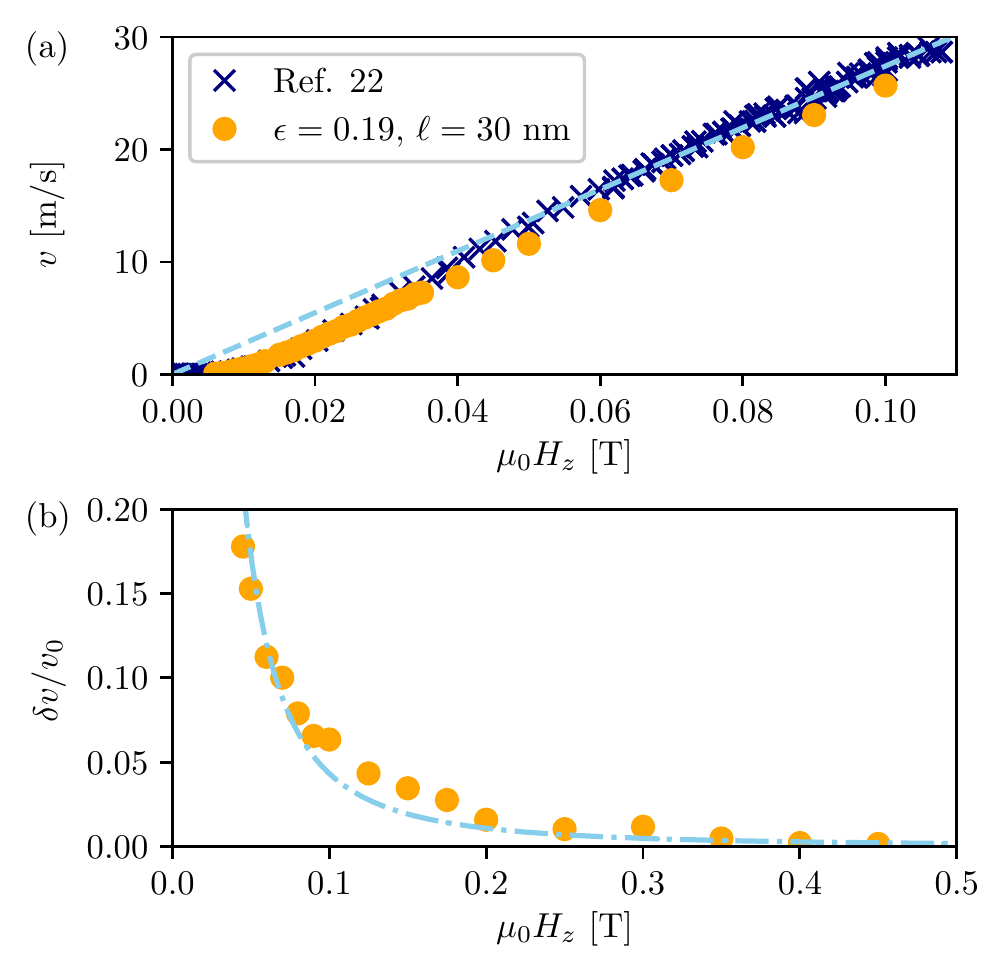}
    \caption{(a)~Values of $v$ vs. $\mu_0 H_z$ in a system with quenched Gaussian Voronoi disorder, compared to the experimental data~\cite{metaxas2007creep}. When disorder is on, the simulated velocity takes really long to reach the flow regime {fitted from the experimental data} (dashed line). (b)~The relative difference between the {expected flow velocity and our numerical results} goes asymptotically to $0$ as $\delta v/v_0\sim 1/(\mu_0 H_z)^{2}$ (dot-dashed line). Note that the two figures have different field scales.}
    \label{fig:flow}
\end{figure}

{Finally, we discuss how fast the flow regime is reached in the {tuned Ginzburg-Landau model}. It has been shown in numerical simulations using the elastic line model~\cite{bustingorry_thermal_rouding_long} and the traditional scalar-field approach~\cite{caballero2018magnetic} that the crossover from the depinning regime to the flow regime is rather slow, as compared to the experimental case~\cite{pardo2017}. The slow approach to the flow regime is also present in our numerical model, as shown in \fref{fig:flow}(a).
As predicted in~\cite{chen1995}, the relative difference $\delta v/v_0=(v_0-v)/v_0$ between the simulated velocity $v$ and the expected value in the flow regime $v_0=m_{\mathrm{f}} \mu_0H_z$, where $m_{\mathrm{f}}$ is the mobility, vanishes asymptotically as $1/(\mu_0 H_z)^{2}$ (\fref{fig:flow}(b)). Therefore, all {these models} seem to share the slow approach to the flow regime, thus still missing some ingredient of the experimental counterpart.}

\section{Domain wall width}\label{sec:width}

\begin{figure}[tb]
    \centering
    \includegraphics[width=0.7\linewidth]{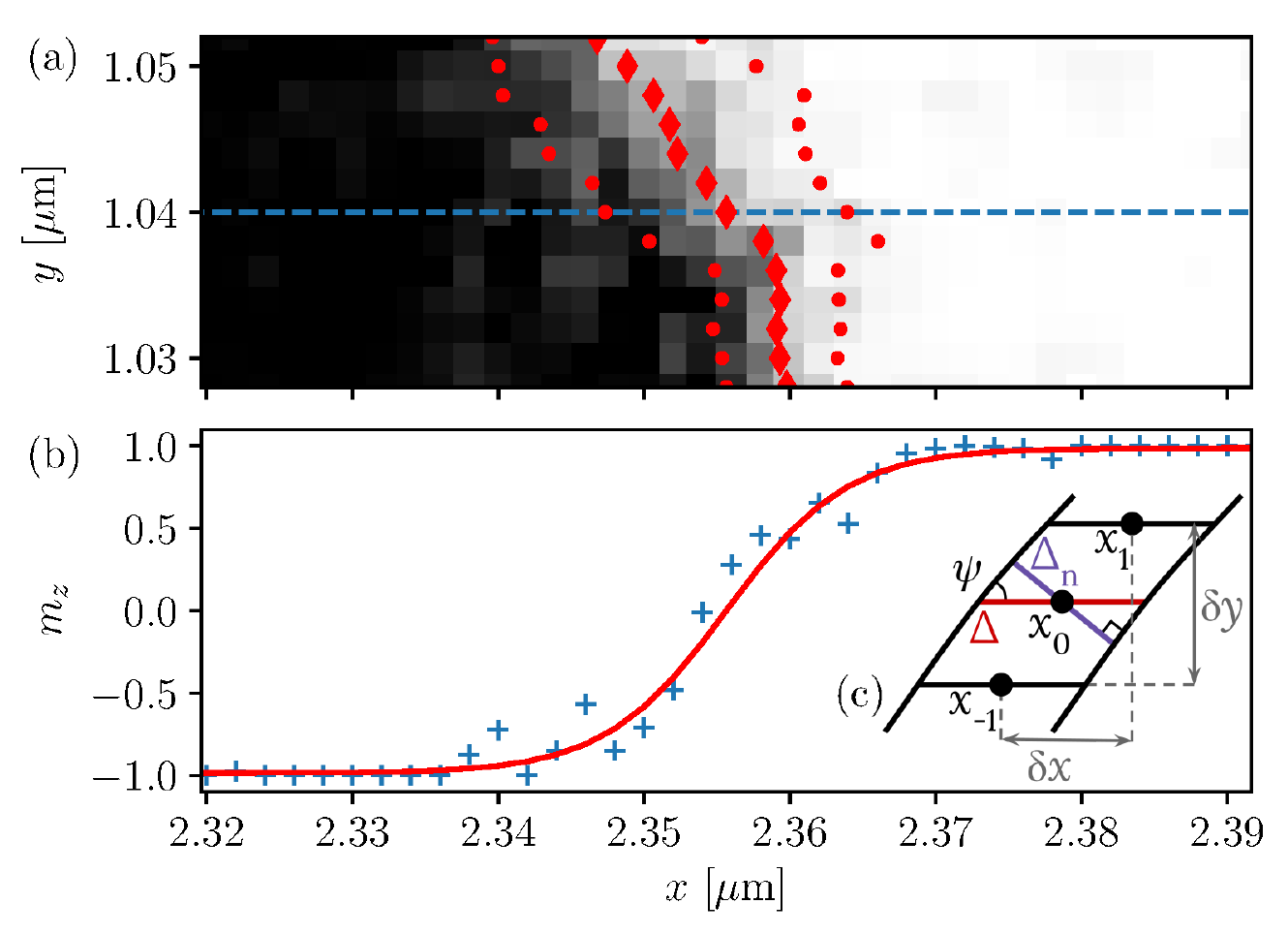}
    \caption{(a)~Zoomed view of the system in the region where $m_z$ changes from $-1$ to $+1$. Gray levels represent the value of $m_z$ in each cell, highlighting the internal structure of the domain wall. The wall is characterized by its position $u(y)=\lbrace x_0(y)\rbrace$ (diamonds) and mean width {parameter} $\langle\Delta\rangle$ (dots standing for $x_0(y)\pm\Delta(y)$). (b)~These values are extracted from the $m_z(x,y=\mathrm{const.})$ profile (crosses) by fitting equation~\eref{mztanh} (full red line). The profile shown corresponds to the dashed blue line above. (c)~Diagram of a tilted piece of the domain wall. The normal-width parameter $\Delta_{\mathrm{n}}(y)$ at $x_0(y)$ can be obtained from $\Delta(y)$ by taking into account that $\tan\psi= \delta y/\delta x$.}
    \label{fig:wall}
\end{figure}

\begin{figure}[tb]
    \centering
    \includegraphics[width=0.7\linewidth]{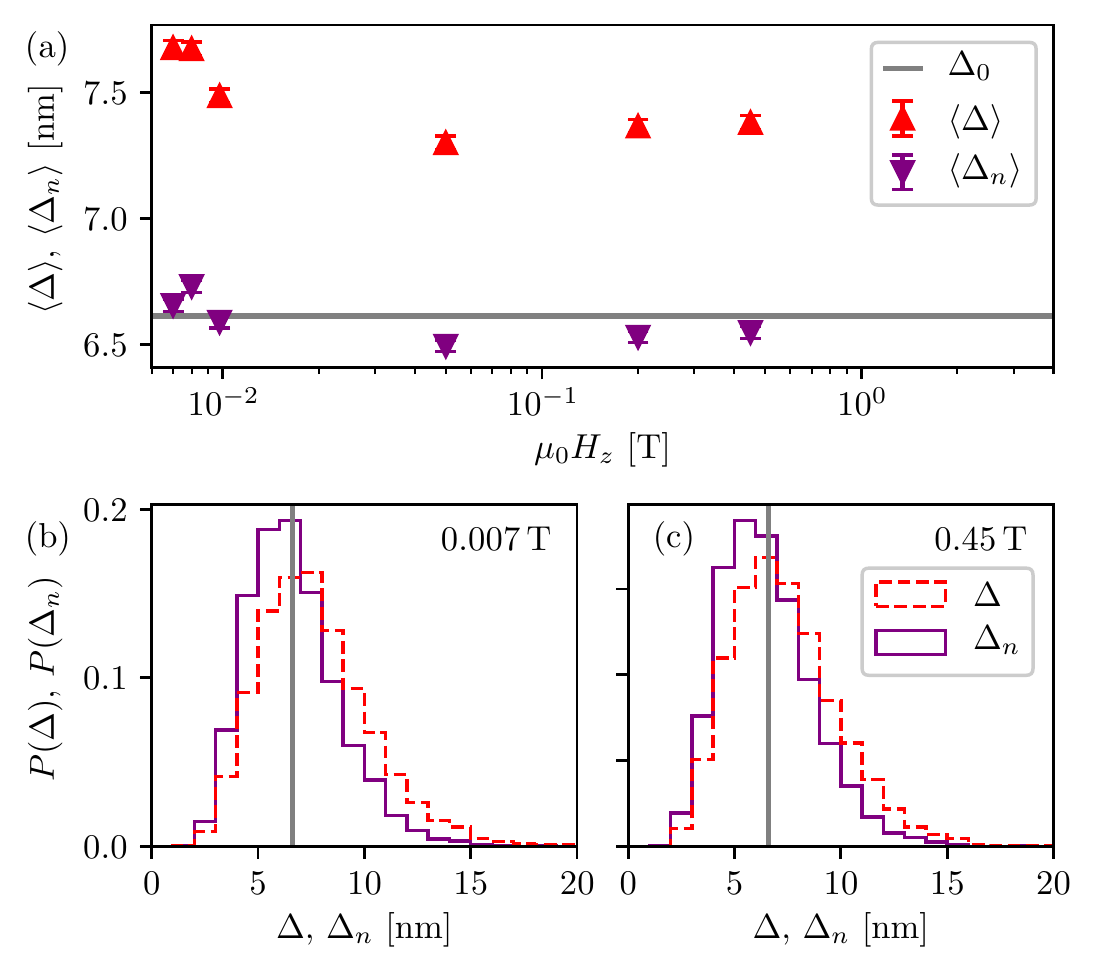}
    \caption{Domain-wall width {parameter} distribution. (a)~The mean value of the width {parameter} defined along the $x$ axis (red) and perpendicularly to the wall (purple), {with error bars representing the corresponding standard errors $\sigma_{\Delta}/\sqrt{2L}$ with $\sigma_{\Delta}$ the standard deviation.} They both present very wide, asymmetrical distributions, as shown with the same color code for (b)~$\mu_0H_z=0.007\,\mathrm{T}$ and (c)~$\mu_0H_z=0.45\,\mathrm{T}$.}
    \label{fig:width}
\end{figure}

While it allows us to simulate systems comparable to experiments, this {tuned Ginzburg-Landau model} provides at the same time access to small length scales beyond PMOKE capabilities. In this {section} we shall focus on the {characteristic shape of magnetization profiles along a domain wall}, which can be seen in \fref{fig:wall}(a).
{We begin by slicing the system in paths parallel to the $x$ axis. As shown in \fref{fig:wall}(b), we then fit the resulting profiles with the expression
\begin{equation}\label{mztanh}
    m_z(x)=\tanh\left(\frac{x-x_0}{\Delta}\right)
\end{equation}
in order to determine the domain wall position $x_0(y)$ and width {parameter} $\Delta(y)$ for each constant $y$}. The position $u(y)$ of the full domain wall is then given by the succession $\lbrace x_0(y)\rbrace$.
The mean value of the width {parameter} measured along the $x$ axis is presented in \fref{fig:width}(a) as a function of the out-of-plane magnetic field. It appears to have a slight dependence on the field, increasing within the creep regime.

There are many ways in which the domain wall width can be defined.
{The previous approach has the drawback of not taking into account the local tilting $\psi$ of the domain wall position $u(y)$}, which can be significant in presence of temperature and disorder. This can explain, for example, the greater value of the width {parameter} at low fields as a consequence of the increased roughness of the wall (see \fref{fig:system2}(a) and \fref{fig:system2}(b), corresponding to fields in the flow and creep regimes, respectively).
Another possibility is, then, to measure the width normal to the wall at each point. 
As shown in \fref{fig:wall}(c), $\Delta_{\mathrm{n}}(y)$ for a given point $x_0(y)$ can be calculated from the width {parameter} measured along the $x$ axis as
\begin{equation}
    \Delta_{\mathrm{n}}(y)=\Delta(y)\sin\left[\arctan\left(\frac{2l}{|x_{-1}-x_{1}|}\right)\right]\, ,
\end{equation}
where we have defined $x_{-1}\equiv x_0(y-l)$ and $x_{1}\equiv x_0(y+l)$ as the positions of the previous and next points in the wall, and $2l$ is their (constant) separation along the $y$ axis. Figure~\ref{fig:width}(a) shows the mean value of the normal-width {parameter} as a function of the field. Now, $\langle\Delta_{\mathrm{n}}\rangle\approx\Delta_0=\sqrt{{A}/{K}}$ in all the field range.

\begin{figure}[tb]
    \centering
    \includegraphics[width=0.7\linewidth]{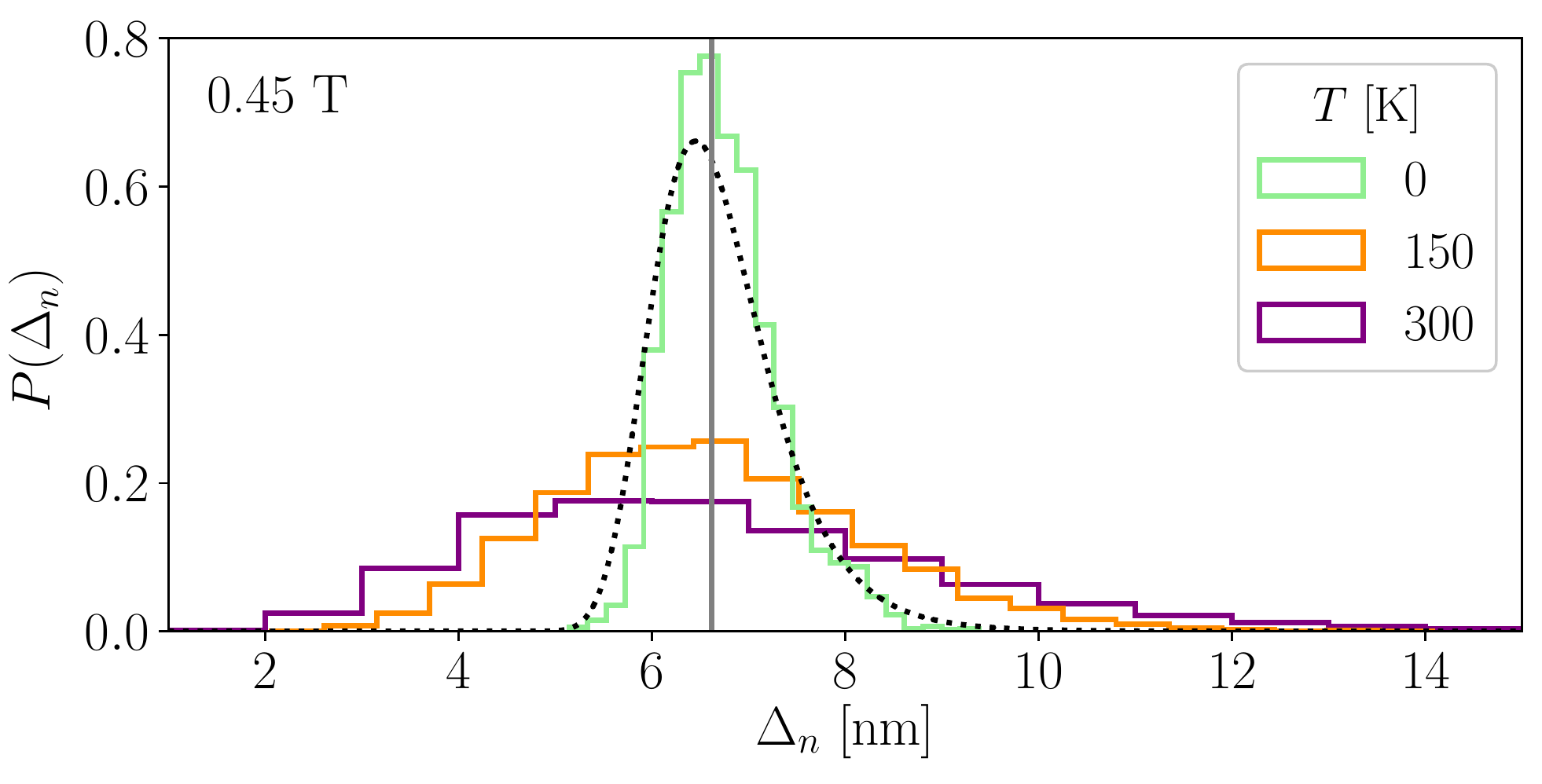}
    \caption{{Probability distributions of the normal-width parameter measured at different temperatures for $\mu_0H_z=0.45\,\mathrm{T}$ (the purple curve is the same as in \fref{fig:width}(c)). A plot of equation~\eref{PDelta} (dotted line) for $\epsilon=0.19$ shows the effect of the quenched disorder and {describes} the distribution found at $T=0$. The vertical line highlights the position of $\Delta_0$.}}
    \label{fig:dist}
\end{figure}

Finally, it is worth mentioning that important fluctuations of the domain wall width {parameter} are present, as observed in \fref{fig:wall}(a). From the values of $\Delta(y)$ and $\Delta_{\mathrm{n}}(y)$ we present in \fref{fig:width}(b) and \fref{fig:width}(c) the normalized histograms corresponding to two values of the field in the creep and flow regimes.
{Two ingredients can in principle be responsible for the width and asymmetry of the distributions: quenched disorder and temperature.
Indeed, a $T=0$, $\epsilon=0$ simulation yields a planar domain wall with a probability distribution for both $\Delta$ and $\Delta_{\mathrm{n}}$ consistent with a Dirac delta centered in $\Delta_0$.
As mentioned in \Sref{sec:validation}, disorder is introduced as a Voronoi tesselation where each grain of mean size $\ell$ has an anisotropy constant given by a Gaussian distribution centered in $k_{\mathrm{eff}}$ with variance $\epsilon^2$. Consequently, when $L\gg\ell\gg\Delta_0$ as in our case, finite values of $\epsilon$ imply a probability distribution for the normal-width parameter\footnote{{A distribution $P(k'_{\mathrm{eff}})$ for the anisotropy random variable $k'_{\mathrm{eff}} =k_{\mathrm{eff}} \left(1 + \epsilon\zeta\right)$ defines a distribution for the normal-width parameter $P(\Delta_{\mathrm{n}}) = |\rmd k'_{\mathrm{eff}} / \rmd\Delta_{\mathrm{n}}|\,P(k'_{\mathrm{eff}})$ via the relation $k'_{\mathrm{eff}}=2\,a_{\mathrm{eff}}/\Delta_{\mathrm{n}}^2$.}} that can be written as
\begin{equation}\label{PDelta}
    P(\Delta_{\mathrm{n}})=\frac{\Delta_0^{2}}{\Delta_{\mathrm{n}}^3} \frac{2}{\sqrt{2\pi\epsilon^2}} \exp\left[ -\frac{1}{2\epsilon^2}\left(\frac{\Delta_0^2}{\Delta_{\mathrm{n}}^2} -1\right)^2\right] \, ,
\end{equation}
where we have used the fact that $\Delta_0=\sqrt{2a_{\mathrm{eff}}/k_{\mathrm{eff}}}$ (see equation~\eref{effectiveAK}).
\Fref{fig:dist} shows a plot of equation~\eref{PDelta} (dotted line) calculated with $\epsilon=0.19$ as used in the simulations.
The excellent agreement between this curve and the normalized histogram obtained for $T=0$ at $\mu_0H_z=0.45\,\mathrm{T}$ shows that quenched disorder alone induces an asymmetrical distribution of the normal-width parameter. 
However, as can be observed comparing with finite temperature data in \fref{fig:dist}, thermal fluctuations are the main responsible for the width of the distributions in \fref{fig:width}(b) and \fref{fig:width}(c), explaining why behaviors so similar are found in such different regimes.} 

{In this way, we have shown {that our model is} capable of providing detailed information on the nano-scale. An important perspective of this work is, then, related to the possibility of exploring the consequences of a finite domain wall width on the small length scale fluctuations}~\cite{agoritsas2010,agoritsas2013static}.

\section{Conclusions}\label{sec:conclusions}

We proposed a local parametrization of the in-plane magnetization in terms of the out-of-plane component that implied {fixing the internal angle of the domain wall}. {This simplification of the full SLLG micromagnetic model allowed us to tune a previous effective scalar-field approach~\cite{caballero2018magnetic},} {making it material-dependent. We used this model to perform simulations of \chem{Pt/Co/Pt} ultra-thin films, whose anisotropy constant, stiffness and saturation magnetization were obtained from the literature~\cite{metaxas2007creep}. The damping constant was found to be within $6\%$ of the reported value by considering the flow regime obtained in previous experiments. The parameters characterizing the disorder of the media, on their turn, were determined by comparing simulations with only one experimental value corresponding to the creep regime. Nonetheless, we robustly} obtained {a velocity-field curve} that quantitatively reproduced experimental data in the \emph{three} regimes (creep, depinning and flow), finding a good correspondence with experimental results for a {large} field range.

At the same time, we used our model to study phenomena at length scales out of experimental reach. We showed that care needs to be taken when the domain wall width is defined, in order to consider contributions of the local tilting. In particular, the theoretical value for the domain wall width {parameter} ${\Delta_0=}\sqrt{A/K}$ was only recovered in mean when the width was measured perpendicularly to the wall.
We also studied the distribution of the domain wall width {parameter}, which we found to be very wide and asymmetrical {mostly due to thermal effects}.

{Starting from the SLLG equation, and considering the energy terms described in the text, we derived a simplified model for the dynamics of the out-of-plane magnetization component in the particular case where the angle $\theta$ in equation~\eref{mxy} is fixed along the domain wall.}
{The link between Ginzburg-Landau type models, inspired in statistical mechanics, and {micromagnetic theory} is thus} given by properly writing the different parameters of the scalar-field approach in terms of important material parameters such as the stiffness, the anisotropy constant and the saturation magnetization. This, together with {the micromagnetic prescription for thermal fluctuations} and the assumption that $\theta$ is a constant, allowed us to reach the low-velocity, creep regime. Quite remarkably, at the same time, it was possible to recover universal features like the creep exponent and other characteristics of the underlying depinning transition.
{{The proposed {tuned Ginzburg-Landau model} allows direct comparison with PMOKE experiments, representing thus a powerful numerical approach to domain wall dynamics in magnetic systems with perpendicular magnetic anisotropy, and related problems.}}

\ack
    We thank R.~D\'iaz Pardo, T.~Giamarchi, E.~Jagla, V.~Lecomte, {F.~Rom\'a, A.~Thiaville, the IDMAG team} {and the PAREDOM collaboration} for useful discussions. {We also thank A.~B.~Kolton and E.~E.~Ferrero for their original contribution to the development of the numerical code.}
    P.~C.~G. and S.~B. wish to thank Universit\'e de Gen\`eve for its hospitality during part of the preparation of this work. {N.~B.~C. acknowledges support from the Federal Commission for Scholarships for Foreign Students for the Swiss Government Excellence Scholarship (ESKAS No. 2018.0636) for the academic year 2018-19. N.~B.~C. also acknowledges support from the Swiss National Science Foundation (FNS/SNF) under Division II, and the IDMAG team of the LPS.} This work was supported by Agencia Nacional de Promoci\'on Cient\'\i fica y Tecnol\'ogica (PICT 2016-0069 and PICT 2017-0906) and Universidad Nacional de Cuyo (grants 06/C561 and M083).


\bibliography{biblio}



\providecommand{\newblock}{}


\end{document}